\documentclass[authoryear]{elsarticle}
\usepackage{natbib}

\usepackage{graphics}
\usepackage{graphicx}
\usepackage{epsfig}
\usepackage{color}           % For color text: \color command

\usepackage{amssymb}

\newcommand{\eg}{\textit{e.g.}}
\newcommand{\ie}{\textit{i.e.}}

\newcommand{\adv}{{\it Adv. Spa. Res. }}

\newcommand{\aap}{{\it Astron. Astrophys. }}

\newcommand{\apj}{{\it Astrophys. J. }}
\newcommand{\apjl}{{\it Astrophys. J. Lett. }}

\newcommand{\jastp}{{\it J. Atmos. Sol. Terr. Phys. }}
\newcommand{\jgr}{{\it J. Geophys. Res. }}

\newcommand{\nat}{{\it Nature }}

\newcommand{\pasj}{{\it Pub. Astron. Soc. Japan }}

\newcommand{\solphys}{{\it Solar Phys. }}

\begin{document}

\begin{frontmatter}



\title{Solar filament eruptions and their physical role in triggering Coronal Mass Ejections}

%\thanksref{footnote1}}
%\thanks[footnote1]{This template can be used for all publications
%in Advances in Space Research.}


 %\author{B. Schmieder\corauthref{cor}}
\author{B. Schmieder, P. D\'emoulin,  G. Aulanier}
\address{Observatoire de Paris, LESIA, 92195 Meudon France}

 %\corauth[cor]{B. Schmieder}
%\thanks[footnote1]{Predictive Science, Inc.,9990 Mesa Rim Road San Diego, CA 92121 U.S.A.}
\ead{brigitte.schmieder@obspm.fr}
% \ead{pascal.demoulin@obspm.fr}
%\ead{etienne.pariat@obspm.fr}
%\ead{tibor.torok@obspm.fr}
%\ead{guillaume.aulanier@obspm.fr}
%url can be given like this
%\ead[url]{http://authors.elsevier.com/locate/latex}

\begin{abstract}
Solar filament eruptions play a crucial role in triggering coronal mass ejections (CMEs). 
More than 80\%  of eruptions lead to a CME. This correlation has been studied extensively  during the past solar cycles and  the last long solar minimum. The statistics made on events occurring during the rising phase of the new solar cycle 24 is in agreement with this finding.  Both filaments and CMEs have been related to twisted magnetic fields. Therefore, nearly all the MHD CME models include a twisted flux tube, called a flux rope.  Either the flux rope is present long before the eruption, or it is built up by reconnection of a sheared arcade from the beginning of the eruption.

 In order to initiate  eruptions, different mechanisms have been proposed:  new emergence of flux, and/or dispersion of the external magnetic field, and/or reconnection of field lines below or above the flux rope.  These mechanisms reduce the  downward magnetic tension  and favor the rise  of the flux rope.  Another mechanism is the kink instability when the configuration is twisted too much.
In this paper we open a forum of discussions revisiting observational and theoretical papers to  understand which mechanisms  trigger the eruption. 
We conclude that  all  the above quoted mechanisms could bring the flux rope to an unstable state.  However, the most efficient mechanism for CMEs is   the loss-of-equilibrium or torus  instability, when the flux rope has reached  an unstable threshold determined by a decay index of the external magnetic field.

\end{abstract}


\end{frontmatter}

\begin{figure}      %%%%%%%%%%%%%%%%%%   FIGURE 1
\centerline{
\includegraphics[width=0.5\textwidth,clip=]{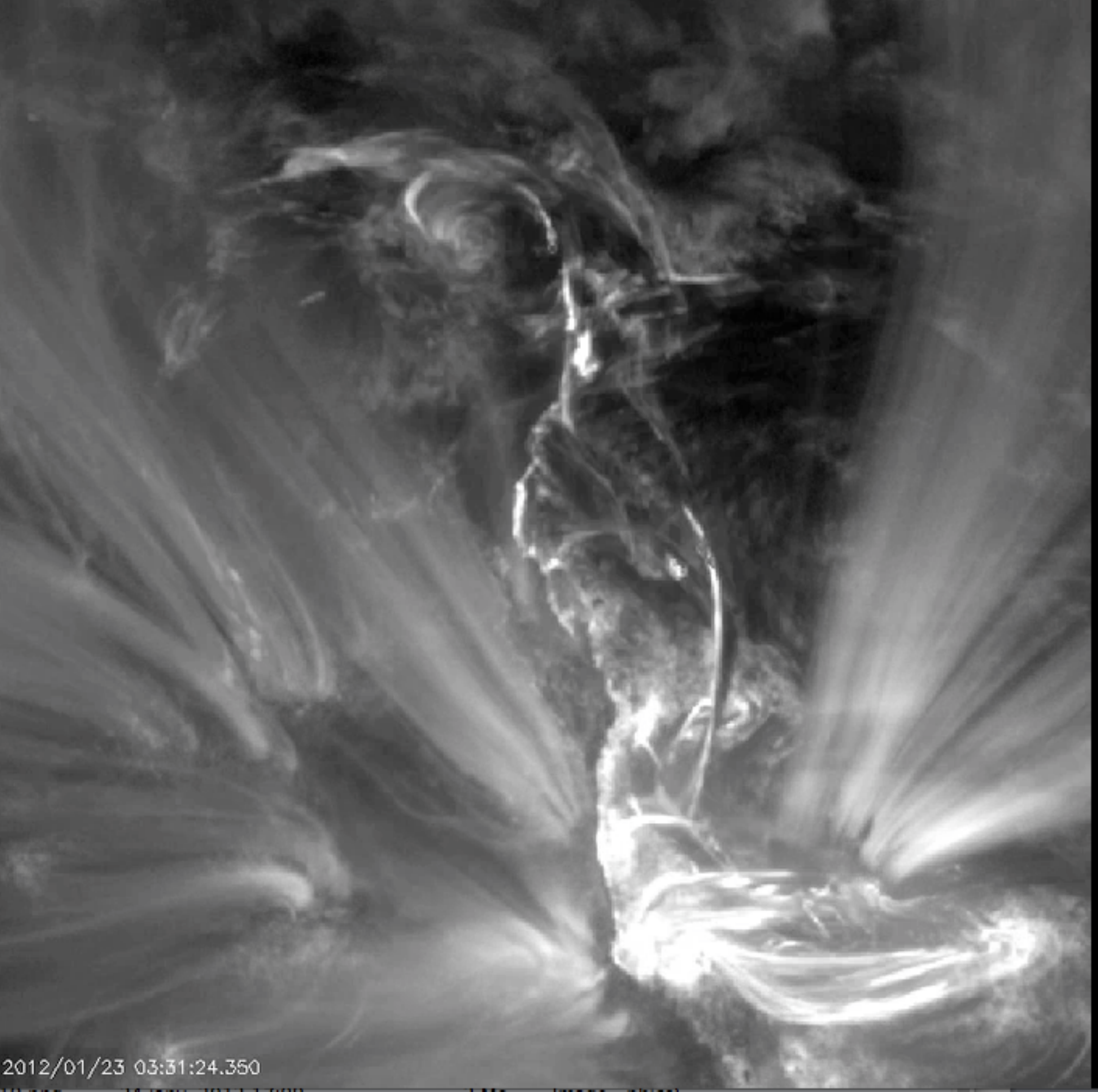}
\includegraphics[width=0.5\textwidth,clip=]{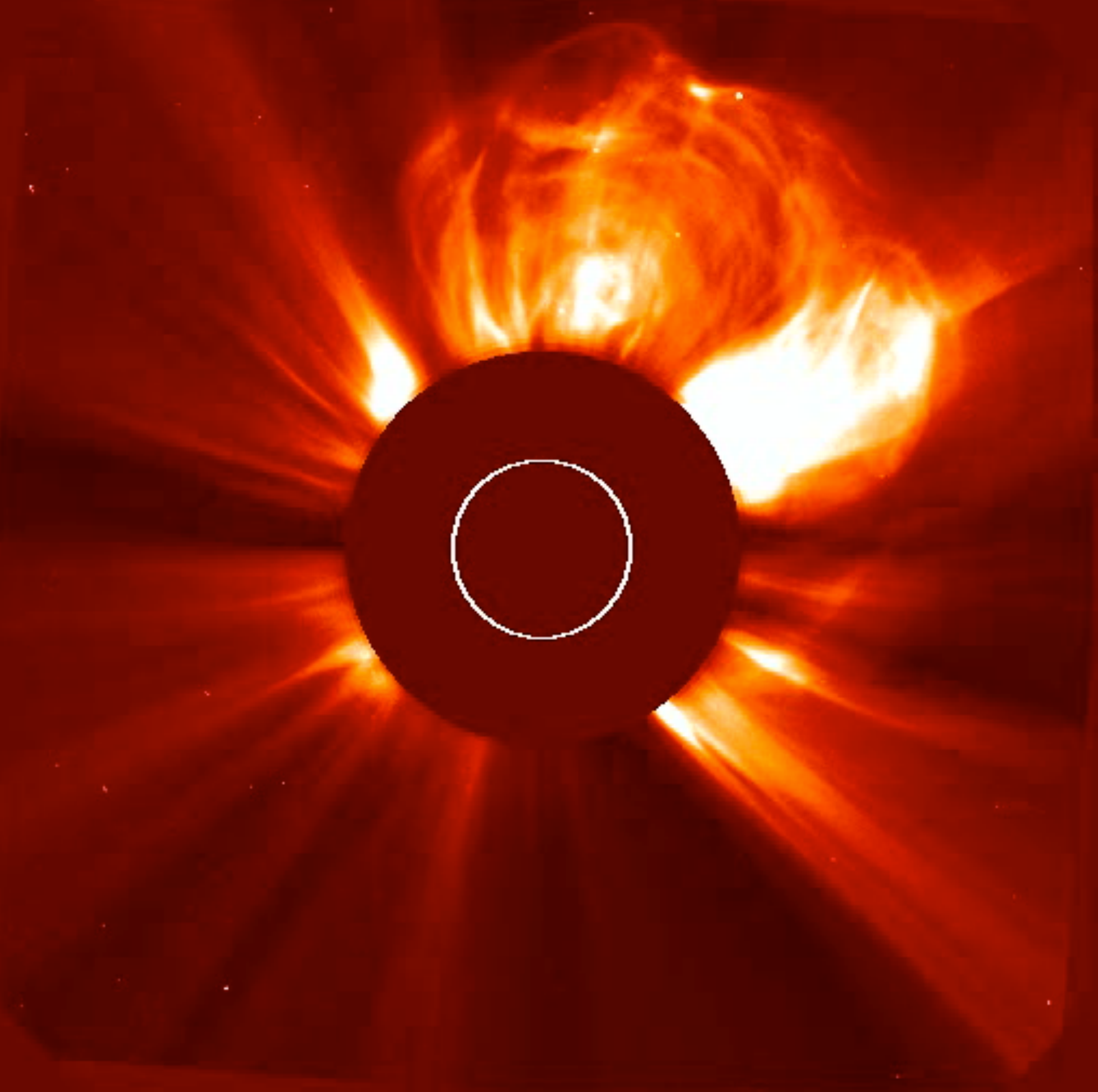}
}
\caption{ {\it Left panel:} Eruption of a filament observed on 23 January 2012 with the 171~\AA\ SDO/AIA filter, just before a M9 class flare at 03:38 UT.  {\it Right panel:} The associated CME observed with SOHO/LASCO C2.  This event was very energetic with a large  amount   of accelerated  protons  (larger than 10 Mev) as registered by GOES.  The 24 solar cycle  was  already very active  at the beginning of 2012  with 7 active regions on the disk \citep{Schmieder12b}.}
\label{january}
\end{figure}

\section{Introduction}
It is widely accepted that solar prominences, with their overlying arcade system, are the progenitors of coronal mass ejections (CMEs) in the corona  (Figure \ref{january}).
Understanding their role in triggering CMEs is a major goal of  solar physics.  The relationship between filament eruption and other active solar phenomena such as flares or CMEs have been extensively  investigated  during the past years \citep[\eg , ][]{Subramanian01,Chandra10}.
Filaments/prominences are located either in active regions (ARs) or between ARs or in quiet Sun (like polar crown prominences). These three types of filaments are all frequently associated with CMEs. 
The rate of the association is very high according to the statistics made for different sets of  events observed during the past solar cycles (56\%, \citealp{Jing04}; 83\%, \citealp{Gopalswamy03};  92\%, \citealp{Hori02}; 62\%, \citealp{Liu12k}). 

\begin{figure}      %%%%%%%%%%%%%%%%%%   FIGURE 2
\centerline{
\includegraphics[width=0.5\textwidth,clip=]{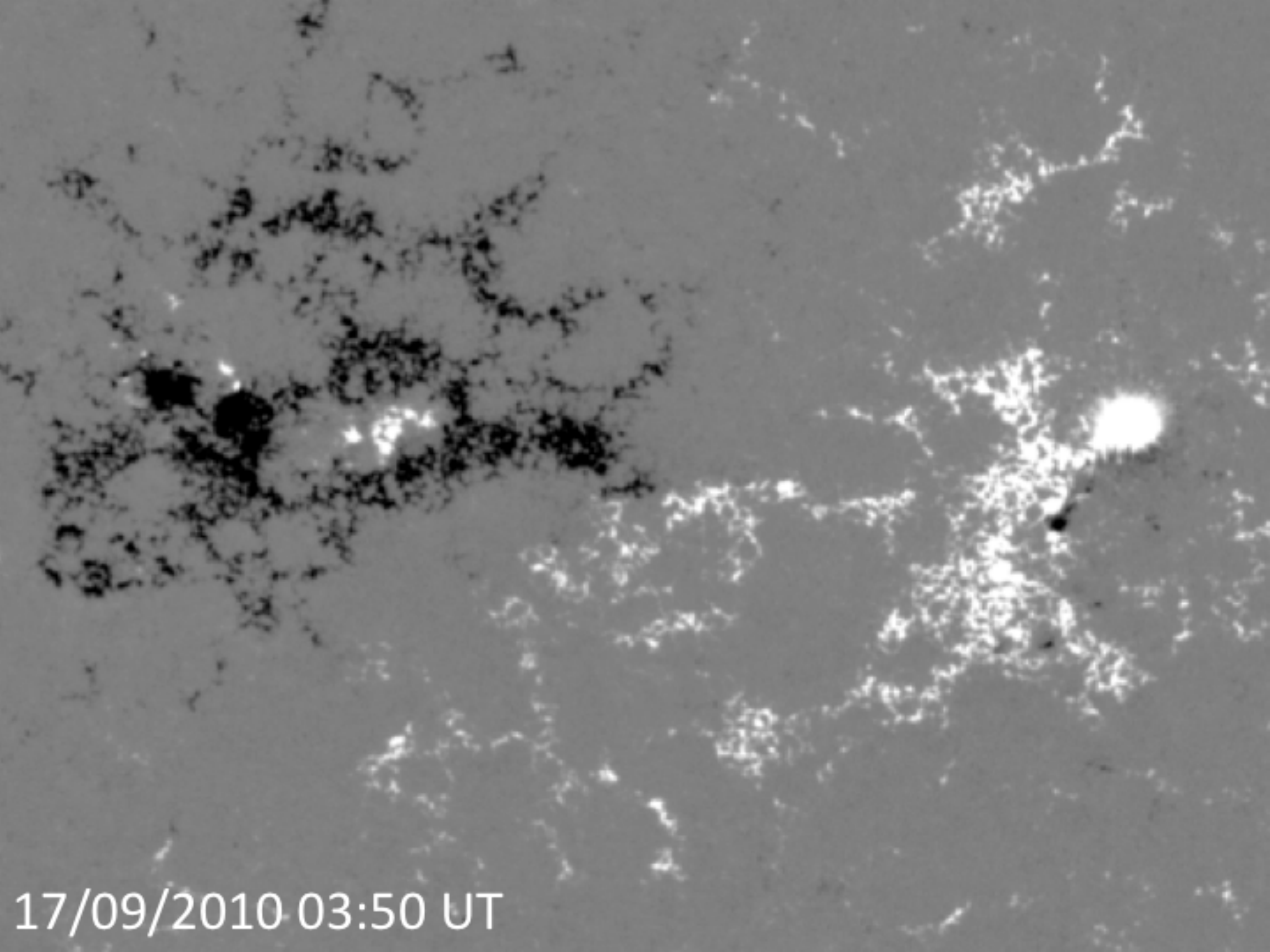}
\includegraphics[width=0.5\textwidth,clip=]{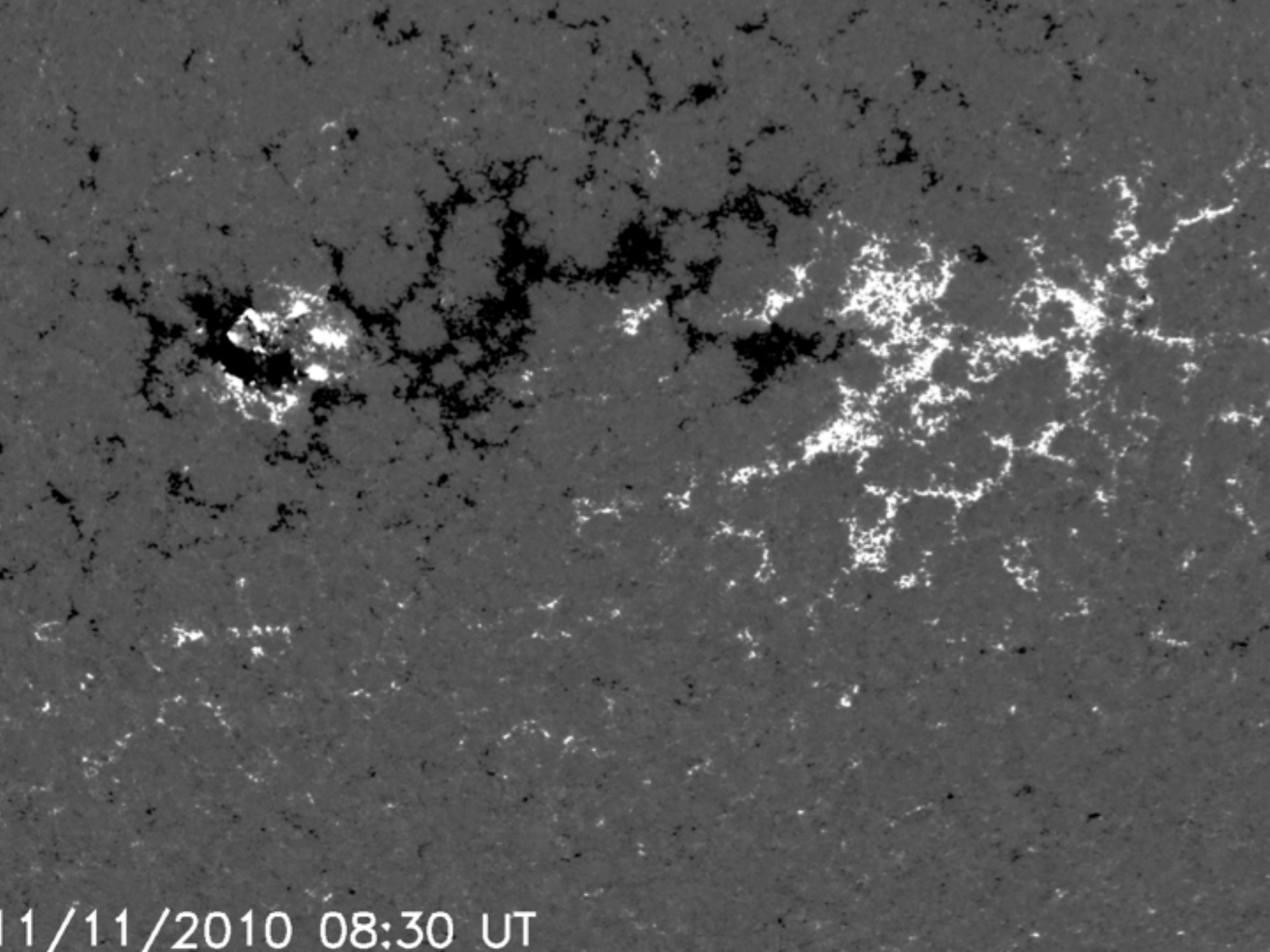}
}
\caption{Emerging flux inserted in the negative following polarity of mature ARs during the rising phase of the solar cycle 24 observed with SDO/HMI magnetograph. {\it Left panel:} AR on 17 September 2010 at 03:50 UT. {\it Right panel:} AR on 11 November 2010 at 08:30 UT.  White/black regions are positive/negative longitudinal magnetic field.}
\label{ars}
\end{figure}

The new solar cycle 24 started by the end of  2010,  after  a long solar minimum lasting nearly two years. Between the beginning of 2010 and  the end of 2011, large long living ARs were observed (\eg, Figure \ref{ars}, Mandrini et al. 2012 in preparation).  The long-lived ARs all had strong magnetic activity with  flares, filament eruptions and CMEs \citep{Schrijver11,Liu12b}. \citet{Li12}  reported that  80 \% of filament eruptions, occurring during the disk passage of a large AR in February 2010, were associated with CMEs. This  is consistent with the statistics of the previous cycle.

Solar flares and CMEs are closely related to the coronal magnetic field.
The plasma  $\beta$ (the ratio between thermal and magnetic pressure) is very small in the low solar corona, and the magnetic energy dominates all
other forms of energies  in the source regions of solar eruptions.  The potential state (current free) magnetic field is the lower bound of energy for a given photospheric vertical field distribution.   Since eruptions require magnetic energy release, the coronal magnetic field must be highly non potential, \ie\ with strong electric currents,  prior to the eruption onset.  
There is much observational evidence in the solar atmosphere  of the presence of such electric currents aligned along the magnetic field lines:  either directly in the photosphere \citep[\eg, ][]{Liu12c}, or indirectly with the presence of J-shaped ribbons \citep[\eg, ][]{Chandra09},  X-ray sigmoids \citep{Green07,MacKenzie08,Liu10,Savcheva12}, twisted filaments/prominences \citep{Williams05,Koleva12}. 

The MHD models of CMEs, commonly include a flux rope. In some of them, the flux rope is already set up in an equilibrium state long before the instability. 
The eruption occurs due to the evolution of the external magnetic field. The process could be: emerging flux \citep{Chen96,Chen00,Jacobs06,Manchester08},  
reconnection of field lines below the flux rope \citep[the tether-cutting model, ][]{Moore92, Moore01}, 
or reconnection above the flux rope \citep[the breakout model, ][]{Antiochos99}.
Another possibility is the presence of an excessively  twisted flux rope \citep[kink instability, ][]{Torok05,Kliem06}. 
Then, many models start with the formation of a flux rope and bring it to an unstable state\ 
 \citep{Amari00,Lin01,Amari05,Forbes06,Fan07,Aulanier10,Olmedo10,Zuccarello12}.
There is also  another  type of CME model assuming a non-flux-rope magnetic structure prior to the eruption.  In this type of  models, the pre-eruption magnetic structure is a sheared core field instead of a flux rope \citep{Antiochos99}. Nevertheless, the sheared core field is converted into a flux rope structure during the eruption through magnetic reconnection. 
As a result, magnetic flux ropes are an important  structural component of CMEs. 

\begin{figure}      %%%%%%%%%%%%%%%%%%   FIGURE 3
%\begin{center}
\centerline{
\includegraphics[width=1.0\textwidth,clip=]{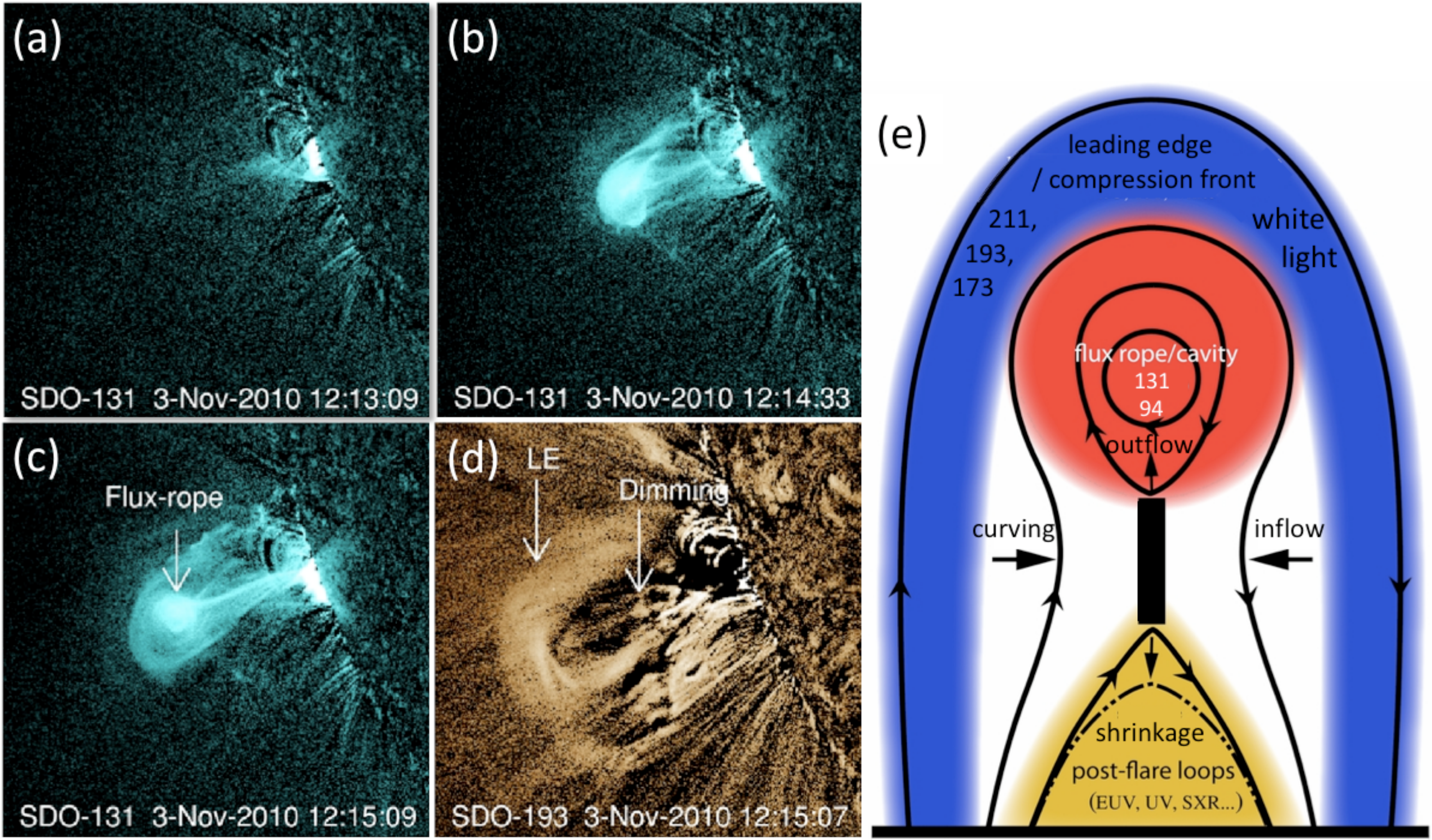}
}
\caption{ AIA 131 \AA\ ($\sim$ 11 MK) base-difference images of the solar eruption on 3 November 2010 (light blue images) and an AIA 211 \AA\  ($\sim$ 2 MK) image (brown image) showing the leading edge (LE) and the dimming. On the right,  schematic drawing of the multi-temperature structures of the solar eruption as it is observed in the low corona by AIA \citep[adapted from][]{Cheng11}.}
\label{cheng}
\end{figure}

\begin{figure}      %%%%%%%%%%%%%%%%%%   FIGURE 4
\centerline{
\includegraphics[width=0.93\textwidth,clip=]{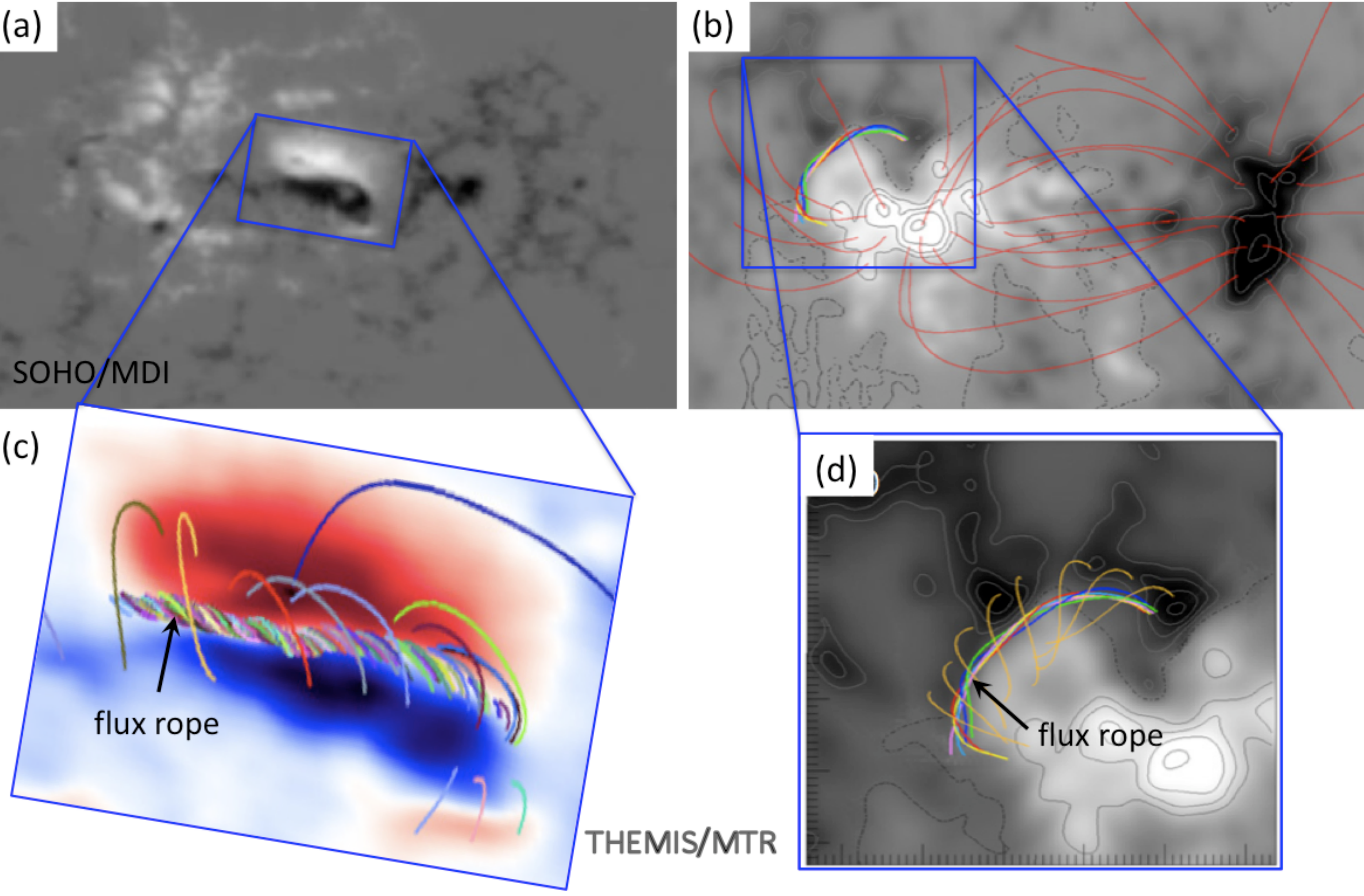}
}
\caption{Non linear force-free magnetic extrapolation of vector magnetograms obtained by THEMIS/MTR showing the presence of a stable flux rope within two ARs. 
(a,b) Magnetograms of the vertical component. 
(c,d) Evidence of the flux rope in the extrapolation of the photospheric magnetograms \citep[adapted from][]{Canou09,Guo10}.
(a) The emerging bipole in the center of an old AR has a "tongue "pattern, which is an indication of the emergence of a flux rope with positive magnetic helicity \citep{Luoni11}.  }
\label{guo1}
 \end{figure}
 
 Filament eruptions are well described by  the  classical 2D flux rope model CSHKP  \citep{Carmichael64,Sturrock66,Hirayama74,Kopp76} and  visualized in limb observations  \citep{Cheng11,Reeves11}. An example is shown in   Figure \ref{cheng}. The observations obtained with 
SDO/AIA filters show the formation and eruption of a flux-rope like structure. The flux rope is  bright in AIA 131 \AA\ images which represents emission of plasma  at  around 11 MK.
In AIA 211 \AA\ (2 MK) stretched loops are passing above the flux rope which is in the dark region (dimming).  At this temperature the leading edge of the CME  is bright due to the enhancement of plasma density by compression in front of the CME.  In the sketch (Figure \ref{cheng}e), based on the classical model CSHKP, the different structures are represented schematically as they appear in the different filters of AIA: the ejection of the hot  flux rope stretching upper  field lines,  the reconnection below the flux rope in a current sheet, and the formation of post-flare loops. These loops become  cooler and cooler as they shrink \citep{Forbes89,Forbes96,Aulanier12}. 
      
 The questions  which arise  from the above discussion are the following:
 \begin{itemize}
 \item 
 Is a twisted flux tube (flux rope) present before the eruption?
  \item
 How  a flux rope is formed above the photosphere?
  \item
 How is  a flux rope  brought to an eruptive state?
 \end{itemize}

   The present review is organized as following. We present in Section~2 a forum that  discussed  whether   or not  a flux rope exists  prior to eruption.  
  In Section~3, we review the eruption mechanisms.  This summary is based on the possible  triggering CMEs processes: emerging flux  or shearing motions in the photosphere.  Their coronal signatures in different temperatures  or wavelengths is also considered.  These  arguments are based on observational as well as theoretical papers. 
  In section~4, we focus on the eruption triggers.  We conclude that the main trigger is the torus instability, which is also a loss of equilibrium.  Moreover, it is argued that the kink instability alone cannot lead to a CME, while in some cases, it can bring the system to a loss of equilibrium.
    These two mechanisms are analyzed in details and we show some examples where the kink instability  alone leads only to confined eruptions.  

\begin{figure}[t]      %%%%%%%%%%%%%%%%%%   FIGURE 5
\centerline{
\includegraphics[width=0.95\textwidth,clip=]{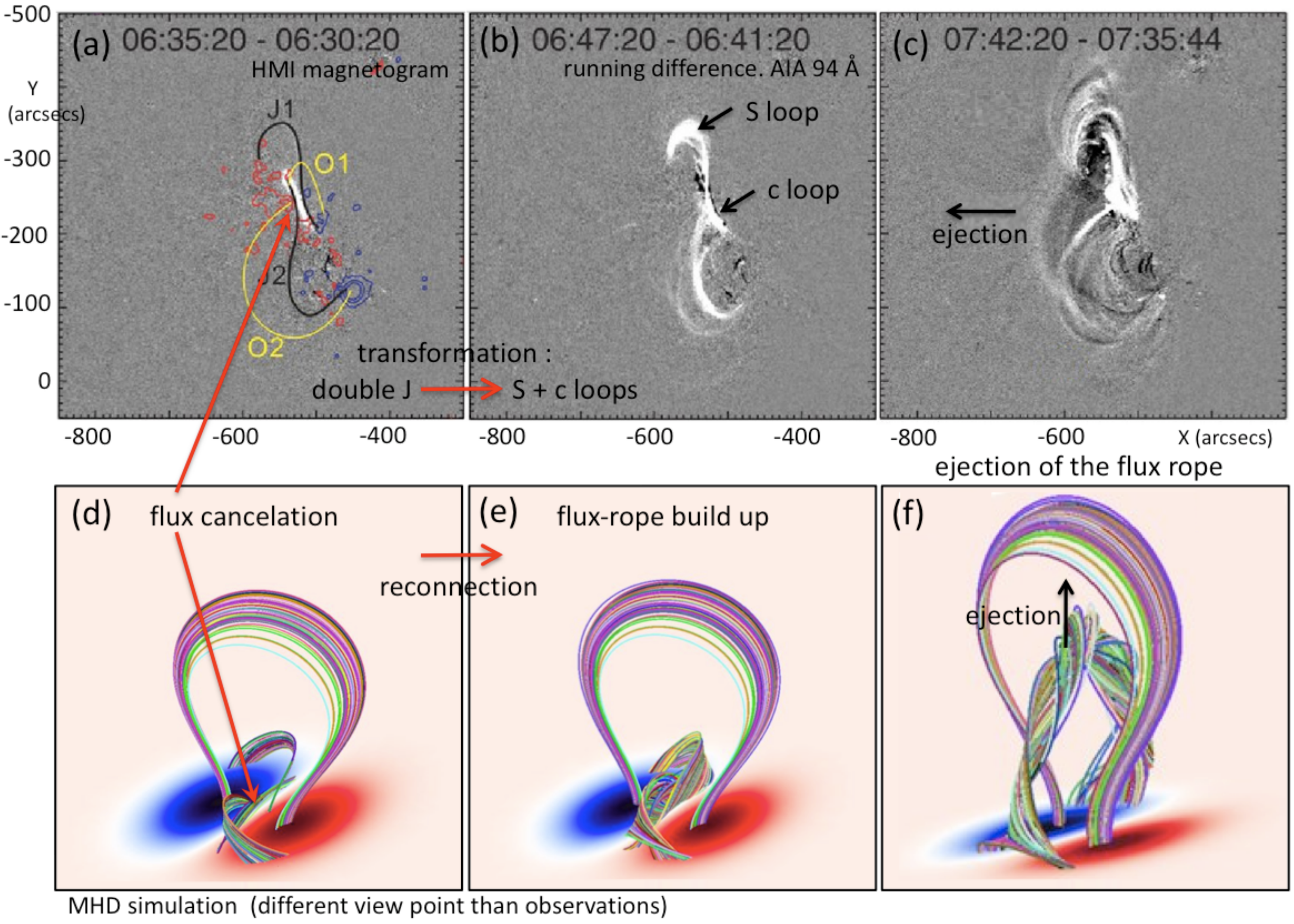}
}
\caption{ Build-up and eruption of a flux rope. (a-c) Formation and eruption of a sigmoid observed by AIA 94 \AA\ near the Eastern solar limb. The presence of a sigmoid is the signature of strong currents aligned with  the magnetic field \citep[adapted from][]{Liu10}. (d-f) MHD simulation showing the built up of a flux rope, by reconnection of field lines due to flux cancelation along the polarity inversion  line and later on the flux rope eruption due to an ideal MHD instability \citep[adapted from][]{Amari10}. The observation and simulation panels have been selected to show a similar phase of the evolution.
}
\label{amari}
 \end{figure}

\section{Presence  of flux ropes } 

\subsection{Evidence of flux ropes} %tube in filament configuration }
 
 Recent observations made with  Hinode/SP (spectropolarimeter) based on   the temporal  rotation of magnetic vectors along an inversion line  in an AR   have been interpreted as   the signature of a flux rope crossing the photosphere \citep{Okamoto08}.  During the emergence,  velocity maps of granules obtained in Fe I 6302 \AA\ have suggested  that the flux rope was rising and the filament channel, represented by longitudinal magnetic field lower than 650~G, was enlarged during its passage through the photosphere \citep{Okamoto09}. The interpretation of these observations  is nevertheless  uncertain and must be tested by simulations (see Section 2.3).
 
 Due to the non linearity of the force-free equation, it is difficult to find analytical solutions for non linear force-free field (NLFFF). Therefore, several numerical methods have been developed, such as the Grad-Rubin \citep{Amari99,Wheatland07,Canou09,Amari10},  the upward integration \citep{Wu99},  the magneto frictional \citep{Valori05,Kusano12,Valori12},  the optimization \citep{Wiegelmann08,Guo10,Guo12b}, and  the boundary element \citep[Green's function like,][]{Yan95} methods.   Figure \ref{guo1} shows two examples where flux rope have been evidenced using NLFFF extrapolations.  \citet{vanBallegooijen04} proposed another method by inserting a flux rope in the magnetic region and led the system to relax using a magneto-frictionnal method. The application of the method to observed cases shows promising results \citep{Savcheva09,Su09,Savcheva12,Su12}. The MHD relaxation approach is used recently in global 3D extrapolation of the full disk magnetic field \citep{Jiang12}. 
 
 Using the theoretical coronal flux rope models by \citet{Titov99} and \citet{Torok03},  it has been shown that NLFFF extrapolation codes are capable of reconstructing significantly twisted flux ropes, as well as topological features of AR magnetic fields \citep{Valori05,Valori10}.
Further more, NLFFF magnetic extrapolations are able to reconstruct a flux rope even in a complex magnetic topology  from observed vector magnetograms \citep[see Figure \ref{guo1},][]{Canou09,Canou10,Guo10}.
 
\subsection{Formation of active regions}
 
 Active regions are formed by flux emergence through the photosphere with the appearance, then separation of the polarities of many magnetic bipoles.  Typically, they drift progressively apart and the polarities are collected in two main polarities revealing the global structure of the underlying flux tube \citep{Zwaan85,Strous96,Pariat04}.

 The two main polarities  have typically some tongue shape  during the  emergence  phase \citep[][see Figure \ref{guo1} top left]{Lopez-Fuentes00,Chandra09,Luoni11}.  These elongated polarities are traces of the azimuthal component of the flux-rope field. By the end of the emergence phase,  typically a few days for an AR with a flux around 10$^{22}$ Mx,  the tongues retract with the agglomeration of their magnetic flux in the corresponding magnetic polarity.
 
Magnetic dips are detected in the photosphere with vector magnetograms. They are loaded with dense material. The emergence can continue only when reconnection is occurring.  It permits the downward escape of the dense plasma. The consequence of reconnection is detected as chromospheric brightenings and plasma flows \citep{Pariat04}.

 \begin{figure}      %%%%%%%%%%%%%%%%%%   FIGURE 6
%\begin{center}
\centerline{
\includegraphics[width=\textwidth,clip=]{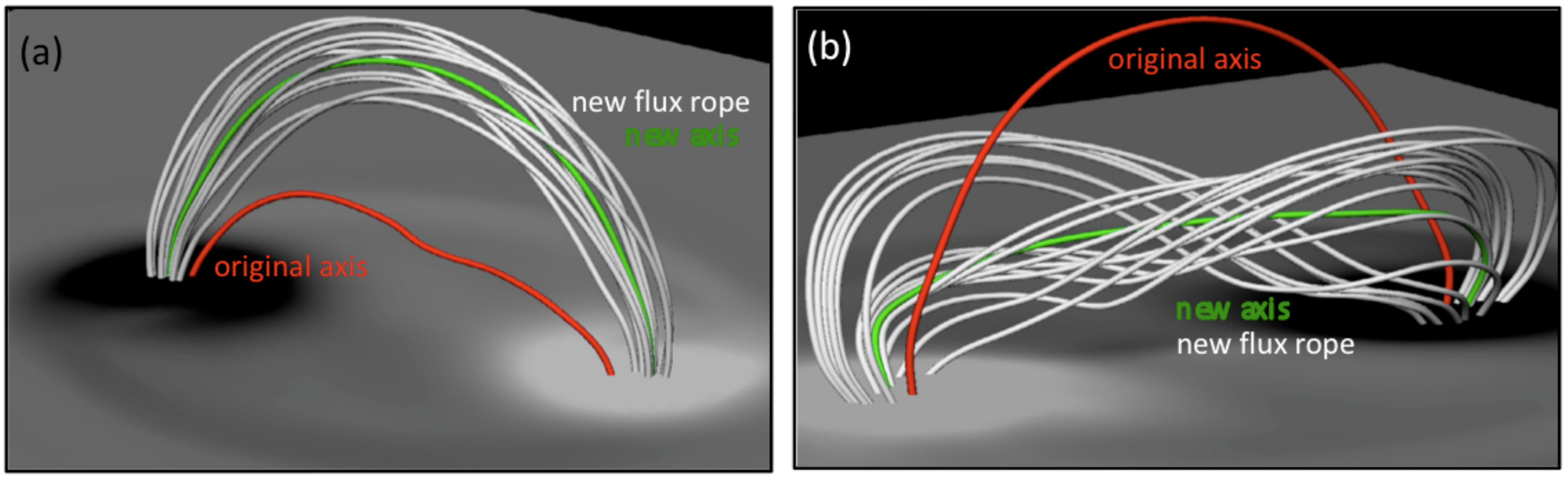}
}
\caption{MHD simulation of the formation of a flux rope above the photosphere for two cases: (a) a weak strength of the magnetic field, (b) a  higher strength of the magnetic field in the simulated flux rope located initially below the photosphere.  In this simulation a flux rope below the photosphere is  partly emerging and there is  a 
 restructuring/reconfiguration of the field above the photosphere in such a way that a new flux rope is created higher than the  original  flux rope axis in case (a) or below in case (b) \citep[adapted from][]{MacTaggart10}.
}
\label{mac}
 \end{figure}

\subsection{Some aspects  of flux rope emergence above the photosphere}
 
 Local MHD simulations calculate the flux rope evolution from below the photosphere to the low corona ($<$10 Mm). The crossing of the photosphere by a flux rope is difficult for several reasons, first the flux rope is no longer buoyant, second, there is a change of regime from high to low $\beta$ plasma, and finally, the flux rope has a much larger radius than the local gravitational scale height, so that its weight   becomes an important downward force which acts against emergence. The flux rope flattens below the photosphere and only fragmented sections  can progressively reach the top of the photosphere \citep{Manchester04,Magara06,Archontis09}. 
 
The above difficulties for emergence were partly solved in the simulations of \citet[][Figure \ref{mac}]{MacTaggart10}.  The rise of the flux rope leads progressively to small reconnections of field lines in the photosphere and finally to a magnetic reconfiguration, as follows. The emergence at and above the photosphere starts with sheared arcades. A pressure depression is present behind the flux rope. This drives converging flows toward the photospheric inversion line, and leads to reconnection of the arcade field lines. This implies 
the formation of a new flux rope above the photosphere. 
Depending on the parametric setting in the initial flux rope (within  the convection zone), the  new flux rope is formed above the previous one or below it  \citep{MacTaggart10}. 
 
Finally, MHD simulations of an emerging flux rope show a patten of magnetic tongues in the photosphere,  similar to the observations of emerging ARs, especially the recent MHD simulations which start  below the photosphere with a flux rope having a curved downward axis \citep{Hood09}.  As in observations, these tongues are present only during the emerging phase when the top part of the flux rope is only partly emerged above the photosphere (\eg, Figure~\ref{mac}b).

 \begin{figure}[t]      %%%%%%%%%%%%%%%%%%   FIGURE 7
\centerline{
\includegraphics[width=0.85\textwidth,clip=]{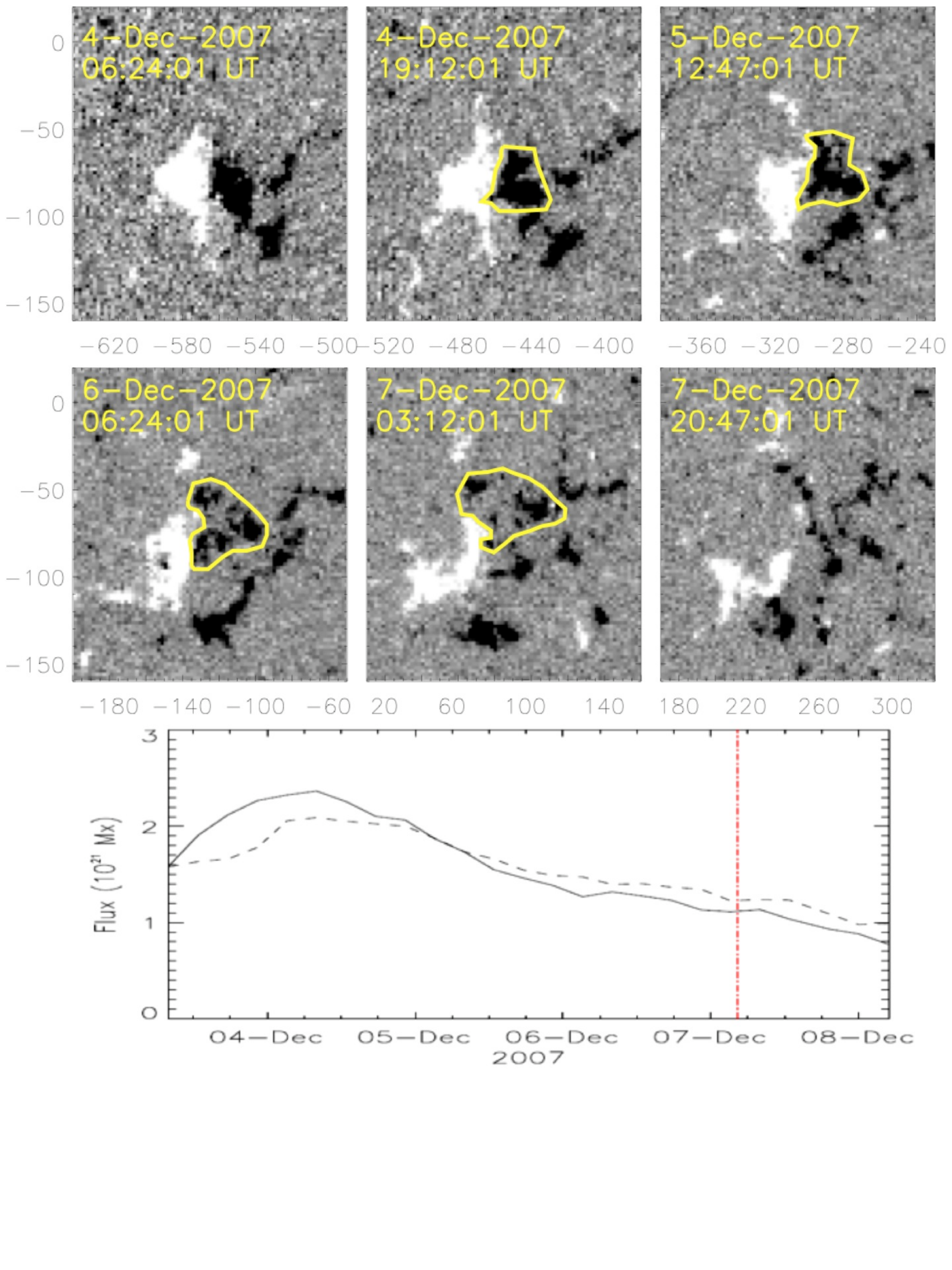}
}
     \vspace{-0.2\textwidth}   % shifted added because white space in the figure 
\caption{Dispersion of the magnetic field in an active  region  producing several  CMEs.  
{\it Top panels:}  MDI  longitudinal derotated  magnetograms of the AR during its disk passage. 
{\it Bottom panel:} Evolution of the magnetic flux of both AR polarities during four days. The red vertical line marks the onset of a CME during the decaying phase of the AR \citep[adapted from][]{Green11}.
}
\label{F_green}
 \end{figure}
  
\begin{figure}[t]      %%%%%%%%%%%%%%%%%%   FIGURE 8
%\begin{center}
\centerline{
\includegraphics[width=0.38\textwidth,clip=]{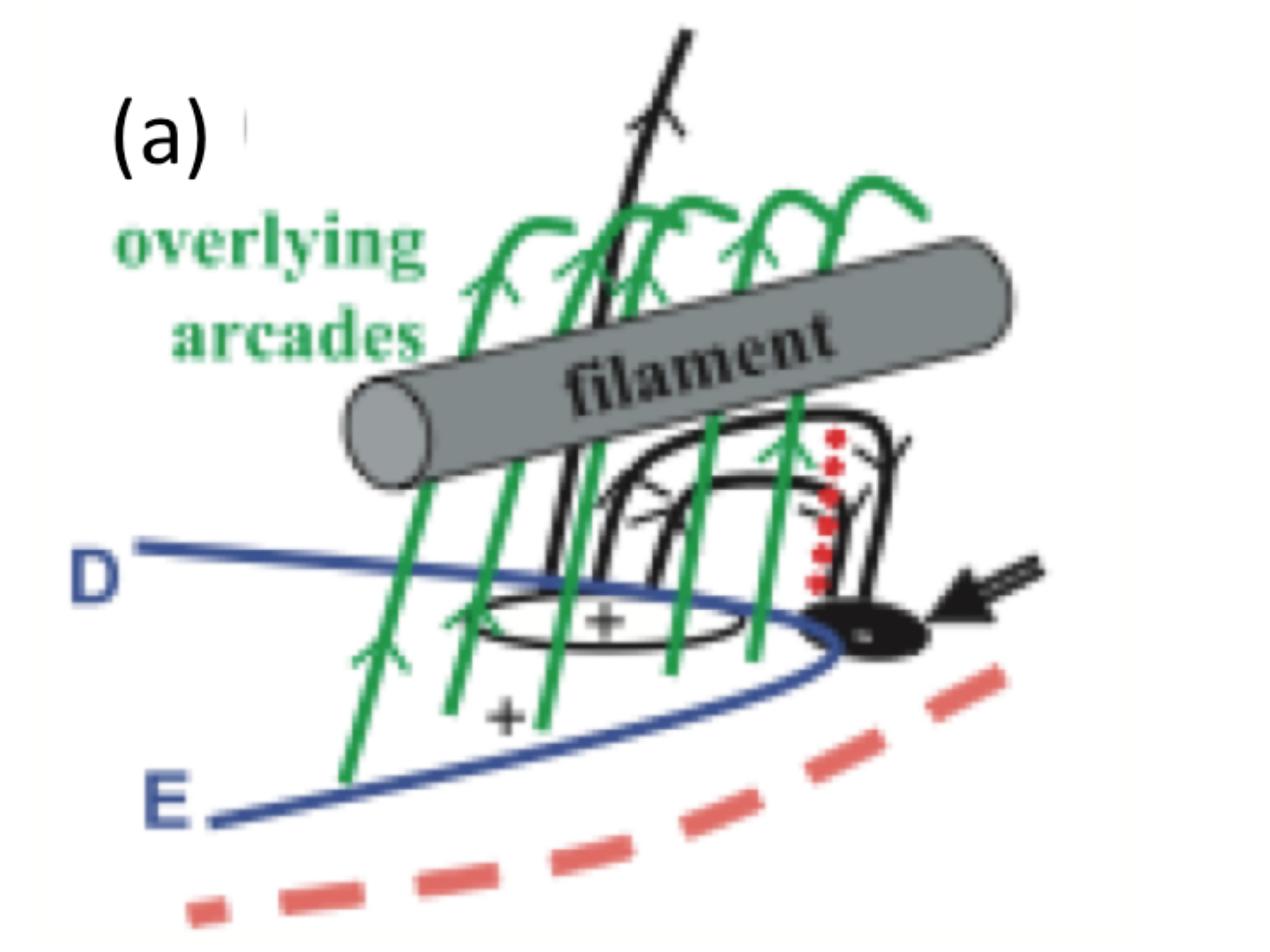}
\hspace{-1cm}
\includegraphics[width=0.38\textwidth,clip=]{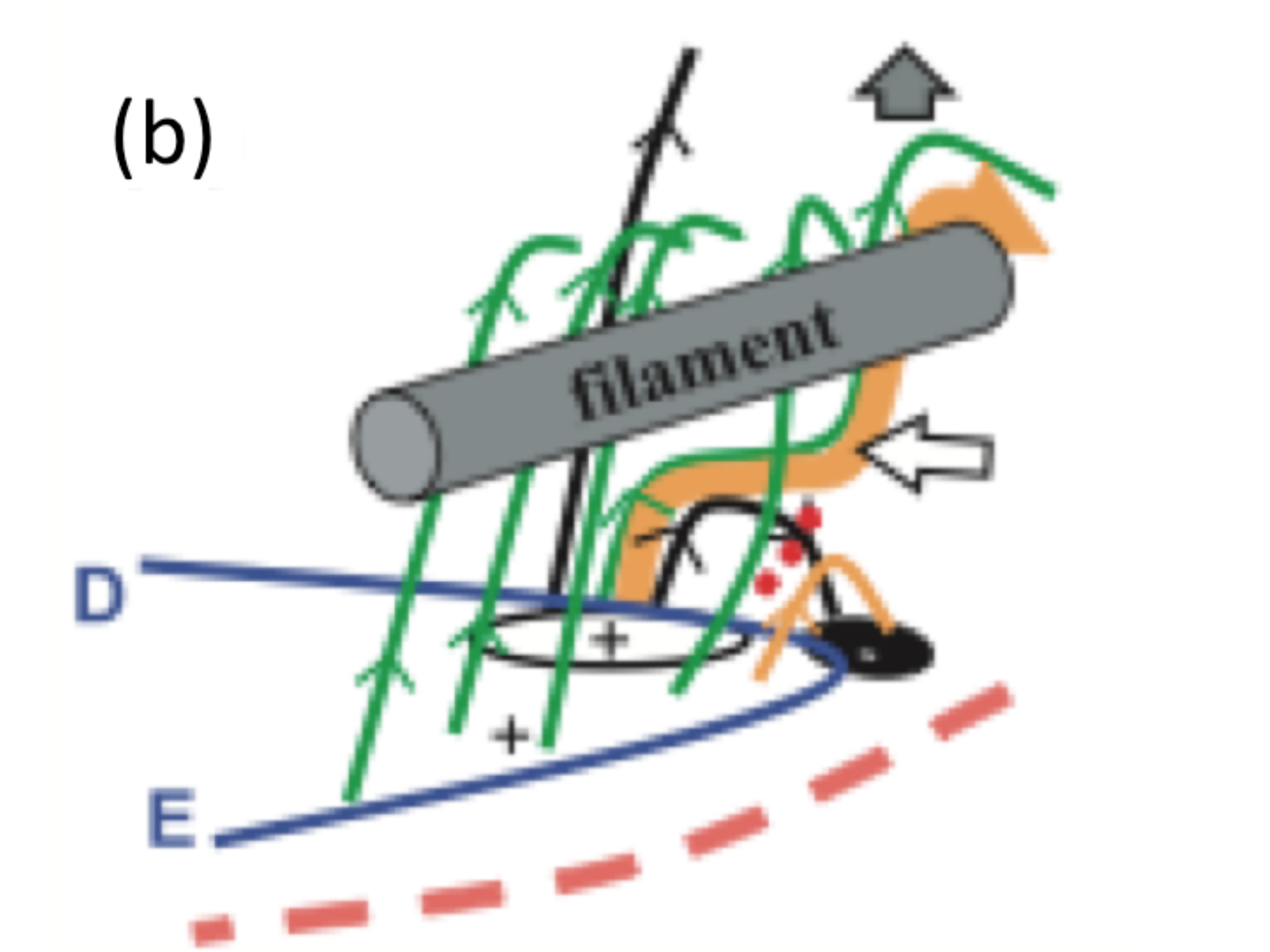}
\hspace{-1cm}
\includegraphics[width=0.38\textwidth,clip=]{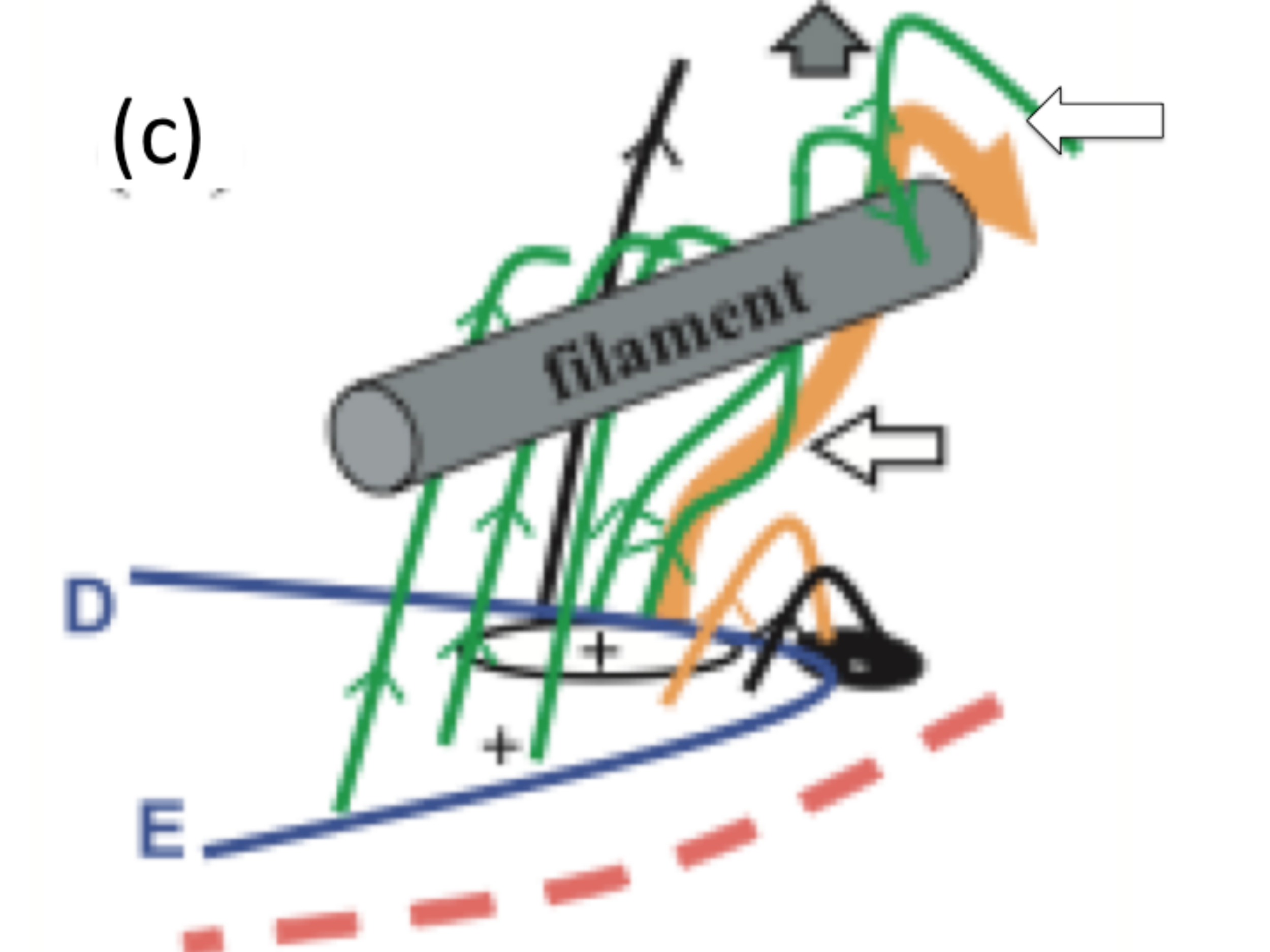}
}
\caption{Schema of the three steps bringing  a flux rope with overlying arcades  in a unstable state at the edge of an AR.  The dotted red line represents its photospheric inversion line.  
(a) Emerging magnetic flux represented by a bipole below the filament.
(b) The emerging field lines reconnect with the arcade field line below the flux rope (indicated by the white arrows) implying a form of tether-cutting.
(c) Above the flux rope, the overlying arcades reconnect with the nearby mature AR in a way similar to the breakout model. 
During the two last phases, the flux rope is rising as indicated by the grey arrows and small X ray flares are observed with Yohkoh \citep[adapted from][]{Nagashima07}.
}
\label{shibata}
 \end{figure}
 
\section{Mechanisms  bringing a flux rope into an unstable state}
 The precise origin of CMEs is still debated  \citep[\eg , ][]{Schmieder12}.  In the above  section we have reviewed how a flux rope can be formed in the low corona.  The presence of a flux tube indicated that the magnetic configuration has a large amount of free magnetic energy that can be released as a CME. 
 What do we need to lift up  the flux rope? The primary necessity is   to decrease the magnetic tension which restrains the flux rope.
There are  two main processes to reduce the tension. First, by progressive reconnection below the flux rope. This is  the tether cutting mechanism proposed by \citet{Moore92} and observed by \citet{Sterling04}. Second, by removing the overlying arcades by coronal reconnection. This is the breakout model proposed by \citet{Antiochos99} and observed by  \citet{Aulanier99}. 
 
Simulations of flux rope eruptions show the importance of the photospheric boundary conditions. \cite{Amari99} modeled a configuration, which can support a prominence based on a flux rope embedded in an overlaying, almost potential, arcade such that high electric currents are confined in the flux rope.  This flux rope is formed by gradual photospheric diffusion of the magnetic field. When this process lasts for a long enough time,  the magnetic configuration cannot stay in equilibrium  and  a CME ensues \citep{Amari00}.  Such photospheric diffusion of magnetic field prior to eruption was  detected observationally by \citet{Schmieder08}.  They observed the  decrease of the magnetic field of  the network,  where the overlying arcades of the filament were anchored,  during two days before the disappearance of a filament using THEMIS data.  The decrease of the total strength in the field-of-view of THEMIS could  be explained by the dispersion of the magnetic field during this time period. \citet{Green11} measured the decrease of the flux in a decaying AR  during the formation of a sigmoid leading to a CME  (Figure \ref{F_green}).
The long term diffusion of an AR was studied by \citet{Demoulin02} and they reported that the rate of CMEs stayed nearly constant in the decaying AR during five months after the emergence phase.

  In fact, in many studies, several  mechanisms seem to work sequentially to bring the flux rope into an unstable state \citep{Sterling07b,Sterling10,Sterling11}.
\citet{Wang93} have  suggested  a two-step magnetic reconnection process: the first step  is a slow reconnection in the lower atmosphere that is observed as flux cancellation, while in the second step, the flare energy release comes from the fast reconnection higher in the corona. 
In the \citet{Nagashima07} paper, the authors explained the observations of an X-ray flare and its associated CME by the presence of a large emerging flux in the decaying AR leading to
a large erupting filament at the edge of the AR. 
The sketch presented in Figure \ref{shibata} summarizes the different observed steps bringing the filament into an unstable state.  Before the eruption there is an emerging flux that is  reconnecting with an arcade overlying the filament. Small flares are also present before the eruption.
 \citet{Nagashima07} pointed out  the fact that these small flares occurred around the footpoints of the large filament but that no eruption was observed. They suggested that  magnetic reconnection at the footpoints of the filament  was not a sufficient condition for  eruption.  The small flares could be the signatures of tether cutting.  Following up, a relatively small (C2.9) flare  was interpreted as reconnection between the flux rope and the AR magnetic field, like in the break-out model.  This last step brought the flux rope into an unstable phase leading finally to the CME and the X1.5 flare in this region.  However, we note that the magnetic configuration needs to be close to the critical point for a loss-of-equilibrium  to trigger the eruption. 
  
 During the solar minimum, many CMEs were initiated by eruptions of large polar crown filaments.  Before eruption, a slow rise of filaments of the order of 1 km/s, during a few hours to one day, have been well observed with the high cadence instruments of STEREO and SDO/AIA \citep{Gosain09,Gosain12,Li12}.  These observations can be directly explained by the loss-of-equilibrium model proposed by \citet{Forbes90} and \citet{Forbes91},  where the flux rope progressively increases in height before erupting.  We suggest that the slow and long-lasting ascending motion of the filament   that is frequently observed corresponds to the change of the equilibrium height of the filament; then the filament approaches a critical point, as described in the loss-of-equilibrium model, and the eruption occurs. 
 
%  Since in the initial condition of many  simulations, the flux rope is already set in an equilibrium state very close to instability or loss of equilibrium, once magnetic flux %emerges, the flux rope can erupt immediately (Chen \& Shibata 2000, Fan and Gibson 2004, Amari et al 2003, 2010, 2011, Jacobs et al 2006, Lin et al 2001).  A loss-of-%equilibrium model was proposed, confirmed by a numerical simulation, that when the filament current exceeds a critical value, the stable configuration containing the %filament loss equilibrium and the filament may erupt (Torok and Kliem 2007). 

 \begin{figure}[t]      %%%%%%%%%%%%%%%%%%   FIGURE 9
%\begin{center}
\centerline{
\includegraphics[width=0.9\textwidth,clip=]{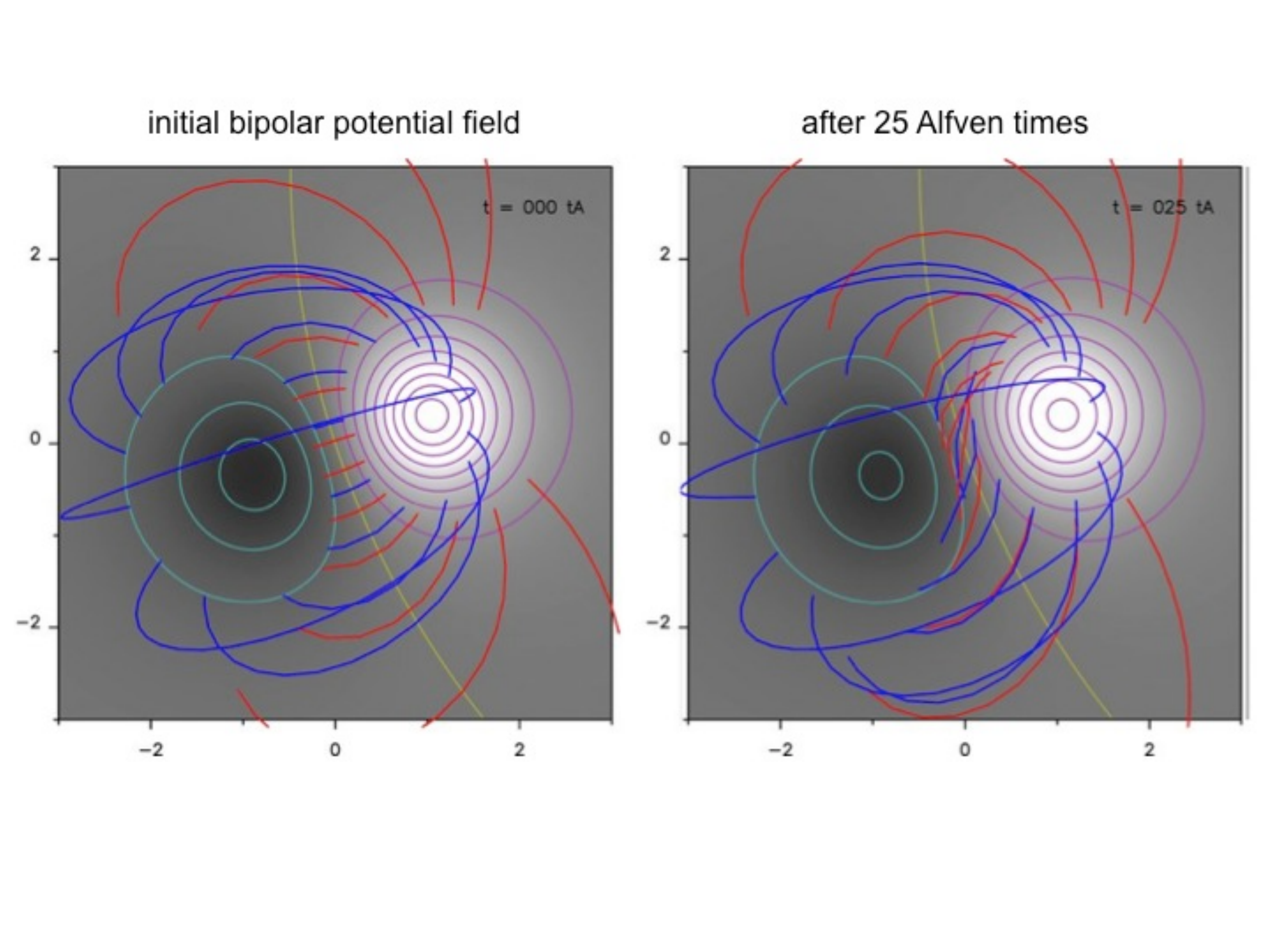}
}
     \vspace{-0.16\textwidth}   % shifted added because white space in the figure 
\caption{ Initial conditions of the model of \citet{Aulanier10} with a bipole  consisting of two polarities with rotation motions at their periphery that creates   a high magnetic shear along the photosphereric inversion line (yellow line).
}
\label{aulanier1}
 \end{figure} 
 \begin{figure}      %%%%%%%%%%%%%%%%%%   FIGURE 10
%\begin{center}
\centerline{
\includegraphics[width=0.9\textwidth,clip=]{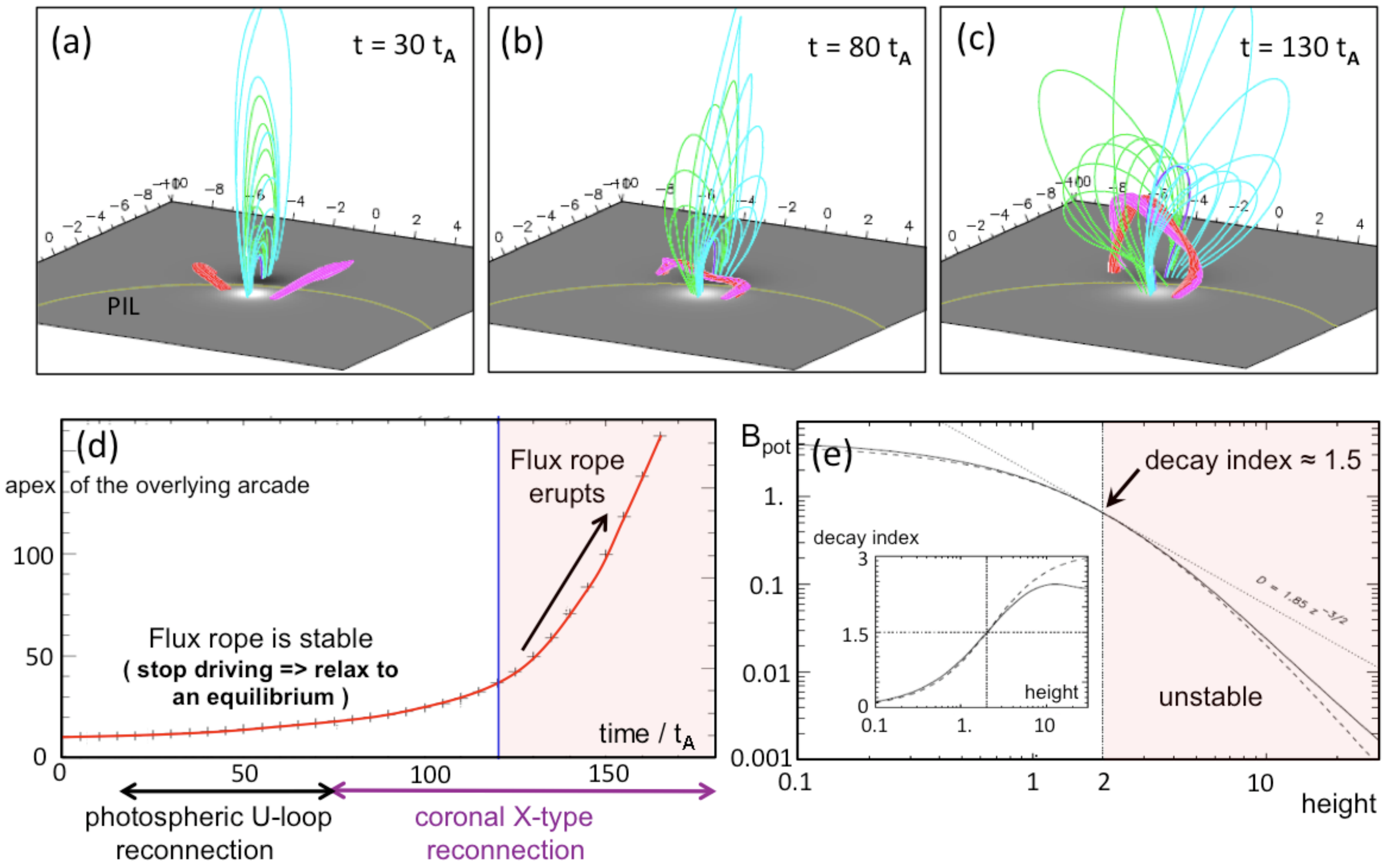}
}
     \vspace{-0.02\textwidth}   % shifted added because white space in the figure 
\caption{ (a-c) MHD simulation of  the formation and  the eruption of a flux rope due to shearing motions and diffusion at the photospheric level.
(a) Initial relaxed potential configuration (as in Figure~\ref{aulanier1}a).
(b) Formation of the flux rope  (pink) after reconnection of the low field lines. 
(c) Beginning of the erupting phase. 
(d) Evolution of the arcade apex height versus time.  
(e) The potential field strength versus height and the decay index showing that for its value $\sim$ 1.5, the flux rope is unstable and can erupt \citep[adapted from][]{Aulanier10}.
}
\label{aulanier2}
 \end{figure}

  \begin{figure}[t]      %%%%%%%%%%%%%%%%%%   FIGURE 11
\centerline{
\includegraphics[width=0.9\textwidth,clip=]{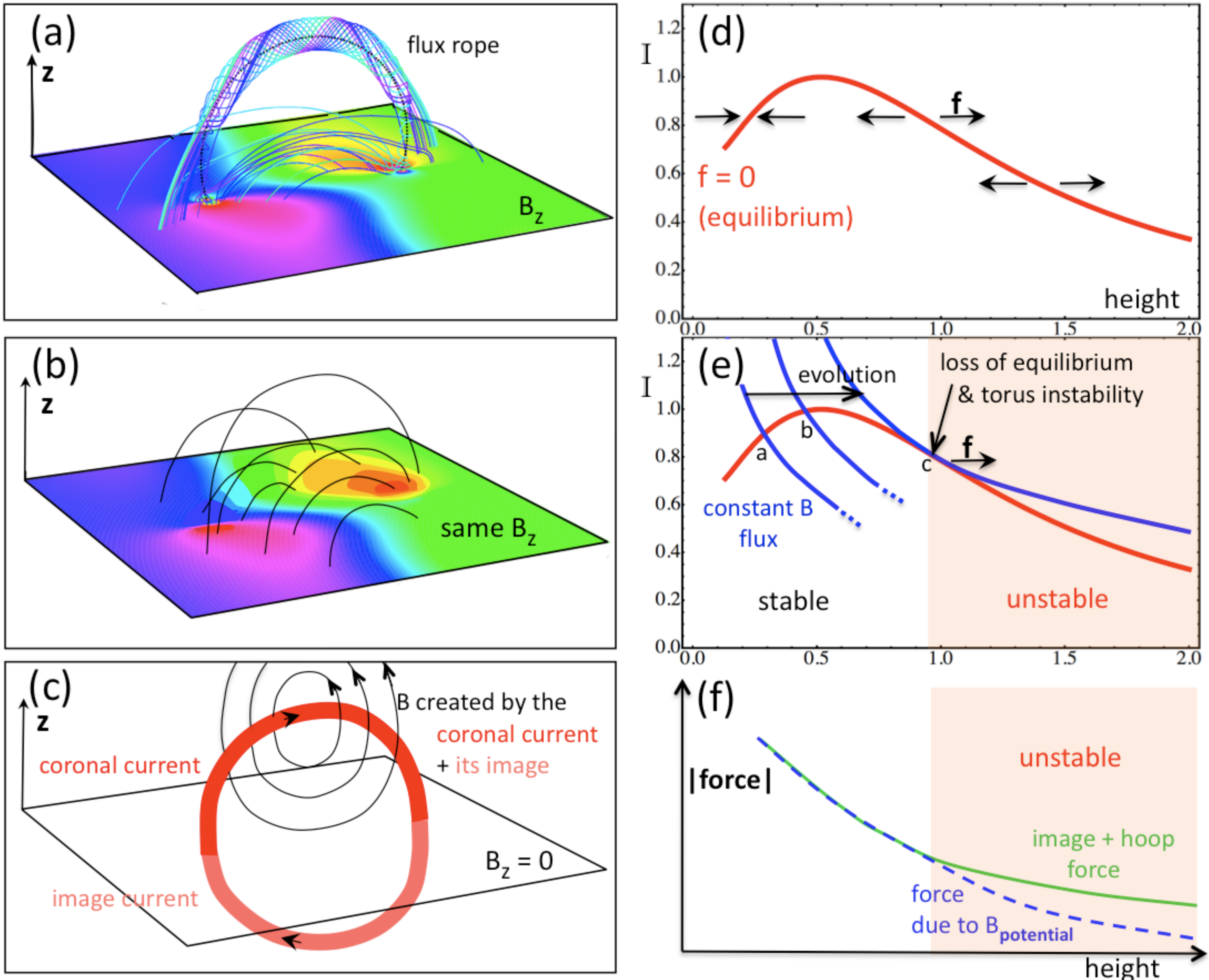}
}

\caption{Analytical demonstration of the equivalence of a loss equilibrium and a torus instability. 
(a) Representation of the flux rope in equilibrium in a bipolar magnetic configuration. This magnetic field is decomposed in the sum of two magnetic field in (b) and (c). 
(b) Potential magnetic field associated to the normal field component  Bz at the boundary. 
(c) Magnetic field created by only the coronal currents and their images (so that Bz=0 at the lower boundary). 
(d) Equilibrium curve (red line) with the current plotted versus height. The black arrows indicate the direction of the force in the vicinity of the equilibrium. 
(e) Evolution with the constraint of magnetic flux conservation until the loss of equilibrium (at point c), which is also the location of the torus instability. 
(f) The forces balance during the equilibrium, while the repulsion of the image current and the hoop force dominate in the unstable region leading to an upward ejection of the flux rope 
\citep[adapted from][]{Isenberg07,Demoulin10}. 
}
\label{instability}
 \end{figure}
 
\section{Trigger mechanisms}
 
 As the magnetic configuration is not directly observed, only simulations, thus far, can test the different  phenomena bringing the flux rope towards eruption.  What is the respective role of emerging flux, twist, shear of the field lines and  the reduced overlying  magnetic field?   \citet{Aulanier10} used their   simulation as  a tool to distinguish the respective role of  these different processes.  The simulation starts with a bipole with rotating sunspots creating a high shear along the photospheric inversion line (Figure \ref{aulanier1}).  Progressively a flux rope is formed by reconnection of  low field lines like in the \citet{vanBallegooijen89} model.  The reconnection at the photospheric inversion line is slowly driven by the applied photospheric diffusion of the magnetic field. As a consequence, the flux rope apex, quasi-statistically, increases its height until it reaches a critical point,  and  then it quickly erupts (Figure \ref{aulanier2}).

\subsection{Analytic model of the torus instability}
 Let us understand, with an analytical analysis, the physical mechanisms which are working  until  the eruption in the \citet{Aulanier10} simulation (Figure \ref{instability}).
The magnetic field of the global configuration can be decomposed in two components, as follows.
The first component is the potential magnetic field created by the photospheric distribution of the vertical field component (magnetogram).  The acting force of the potential field is a restraining force directed downward.  The second component of the field is the magnetic field created by the coronal net current and its subphotospheric image. This induces a magnetic force directed outward, known as the  hoop force (it includes the repulsive force of the image current).  

\citet{Aulanier10}  show that between t=100 and 120 $t_{\rm A}$ (Alfv\'en crossing time of the configuration), the system behavior is similar to the prediction of an electric circuit model (Figure \ref{aulanier2}). The flux rope approaches a critical point of the equilibrium curve driven by a constant increase of the twist (or by changing the magnetic flux below the flux rope).  Their numerical model verifies the non equilibrium conditions analytically calculated with incomplete physics by \citet{vanTend78}, \citet{Bateman78} and \citet{Kliem06}.

 The critical  point  is estimated by the computation of a decay index, which represents the potential magnetic field drop-off along the vertical direction (z-axis):
 $$  n=-d \ln B/ d \ln z$$ 
  \citet{Torok07} found that the decay of the background magnetic field with height is a critical factor in determining whether the instability of the flux rope can result in an eruption or not, \ie , the decay index must be larger than a critical value in order to have a successful eruption. 
The conditions are in accordance with the onset criteria expansion of the "torus instability" which can drive the free radial expansion of an electric current ring in axisymmetric circuit models (Figure \ref{instability}).
 The instability occurs when the decrease in altitude of the downward magnetic tension becomes faster than of the upward magnetic pressure gradient.

  \begin{figure}[t]      %%%%%%%%%%%%%%%%%%   FIGURE 12
\centerline{
\includegraphics[width=\textwidth,clip=]{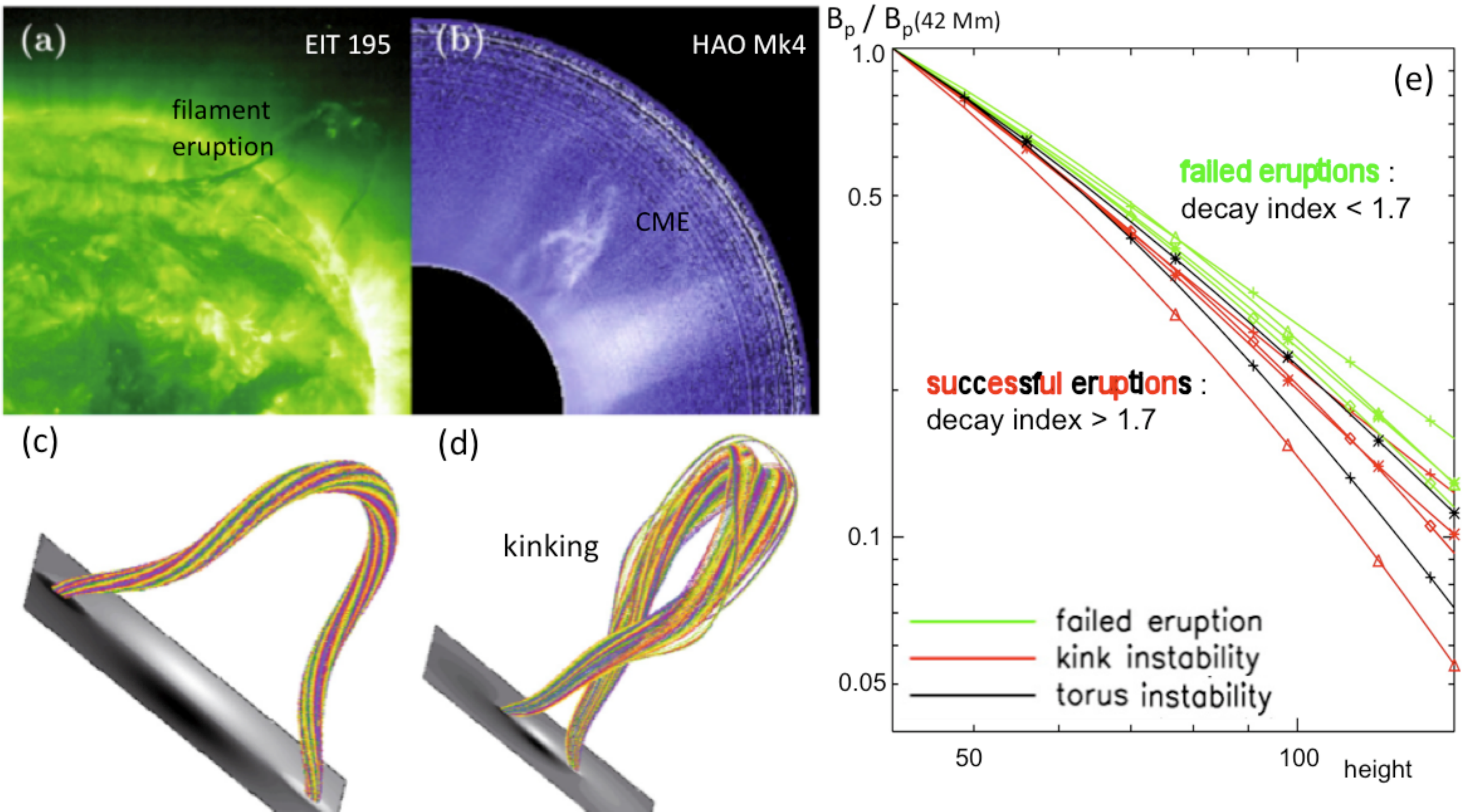}
}
\caption{ Kink and torus instability instabilities. 
(a) Eruption of a filament as seen in EUV. 
(b) The same filament, observed 22 minutes later by the coronograph of HAO, as a kink shaped CME. 
(c,d) MHD simulation with an initial kink unstable flux rope with later the development of some inflation leading to a torus instability. 
(e) The potential field strength versus height for a set of observed eruptions. The decay index of successful eruptions is larger than 1.7, a  value comparable with the model prediction \citep[adapted from][]{Torok10,Liu08}.  
}
\label{kink}
 \end{figure}
 
  \begin{figure}[t]      %%%%%%%%%%%%%%%%%%   FIGURE 13
\centerline{
\includegraphics[width=0.8\textwidth,clip=]{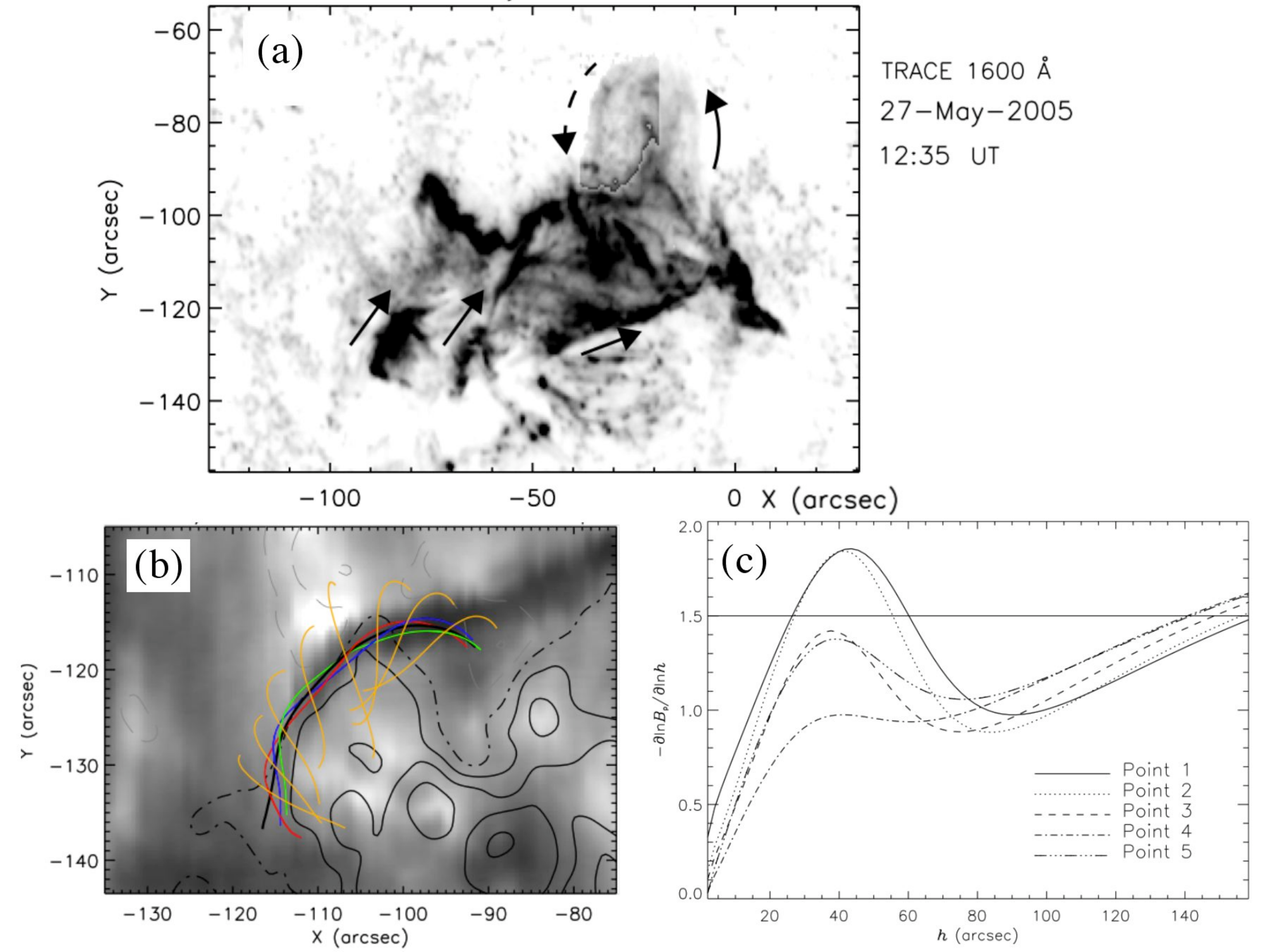}
}
     \vspace{-0.02\textwidth}   % shifted added because white space in the figure 
\caption{ (a) Observation of a filament  eruption   using TRACE 1600 \AA\  filter (reversed color table). The dark lanes  are the bright ribbons, the faint lanes between the ribbons are the flare loops, the faint grey structure  is  the rising filament showing a kink shape. 
(b) Computed field lines, from a non-linear force-free extrapolation. They are drawn on top of an H$_{\alpha}$ image, showing a filament, and isocontours of the vertical component of the magnetic field (see Figure \ref{guo1}b,d). 
(c) The decay index versus height does not reach 1.5. It is the reason why the eruption felt \citep[adapted from][]{Guo10}.
}
\label{guo2}
 \end{figure}

\subsection{Kink instability}
 The kink instability occurs when a flux rope is twisted above a threshold which depends on the twist profile and the aspect ratio of the flux rope.  The kink instability typically leads to a flux rope eruption which saturates towards a certain height  \citep{vrsnak08,Liu12k}  and  could correspond to a  failed eruption if no loss-of-equilibrium  height  was reached (Figure \ref{kink}).
Furthermore, numerical simulations  demonstrate that the helical deformation enforces reconnection between legs of the flux rope if the initial twist is large enough \citep{Kliem10,Karlicky10}. Such a reconnection is complex since it involves also the magnetic field outside the flux rope.
  
  \citet{Liu08} studied 10 events from different ARs, consisting of four failed eruptions, four eruptions due to kink instability, and two eruptions due to torus instability. They calculated the decay index of the background transverse magnetic field in the source AR and found that the decay index for successful eruptions is larger than the one for failed eruptions (Figure \ref{kink}).
\citet{Guo10b}  studied the decay index distribution with height of one confined eruption and found that the decay index is persistently smaller than 1.5 at a height ranging from 5 to 100 Mm above the photosphere.  The magnetic configuration became kink unstable, and  it started to rise at a height of 20 Mm. As a consequence of the low decay index,  and thus the absence of the torus instability,  the erupting filament did not evolve into a CME (Figure \ref{guo2}).

\section{Conclusion}
 
Many observations and simulations indicate that CMEs start with a flux rope eruption.  
Several mechanisms have been proposed involving flux emergence, or/and tether-cutting of field lines below the flux rope or/and photospheric diffusion of the magnetic field.
However, we argue that all the above mechanisms are likely not sufficient to destabilize and eject the flux rope, so as to trigger a CME.
 Several observations and numerical models indeed show that all these effects contribute in building a flux rope from sheared arcades.  Then, slowly, the flux rope is lifted up in altitude.
   In addition, it is worth noticing that many examples show that the kink instability alone only leads to confined eruptions and cannot produce a CME.  Actually, there is no simulation, to
date, which produces a CME with the kink instability working alone.
All previous mechanisms either  bring the flux rope  slowly (emergence, tether-cutting, diffusion), or  rapidly  (kink instability) to a loss of equilibrium, or equivalently a torus instability. This process leads in turn to the ejection of the flux rope and the creation of a CME.

% \bibliographystyle{elsarticle-harv}
% %\bibliographystyle{elsarticle-num-names}
% %\bibliographystyle{elsarticle-num}
% \bibliography{initiationCME}

\end{document}